\documentclass[12pt]{iopart}
\usepackage{listings}
\usepackage{subfig}
\usepackage{hyperref}
\usepackage{graphicx}
\usepackage[utf8]{inputenc}
\usepackage{tikz}
\usepackage[labelfont=bf]{caption}
\usepackage{subfig}


\begin{document}

\title{Evolution of ROOT package management}

\author{Oksana Shadura [1],  Brian Paul Bockelman  [2],  Vassil Vassilev [3]}
\address{[1] University of Nebraska-Lincoln, USA, [2] Morgridge Institute for Research, USA, [3] Princeton University, USA}
\ead{[1] oksana.shadura@cern.ch, [2] bbockelman@morgridge.org, [3] vvasilev@cern.ch}
\vspace{10pt}

\begin{abstract}
ROOT is a large code base with a complex set of build-time dependencies; there is a significant difference in compilation time between the “core” of ROOT and the full-fledged deployment. We present results on a “delayed build” for internal ROOT packages and external packages. This gives the ability to offer a “lightweight” core of ROOT, later extended by building additional modules to extend the functionality of ROOT. As a part of this work, we have improved the separation of ROOT code into distinct modules and packages with minimal dependencies. This approach gives users better flexibility and the possibility to combine various build features without rebuilding from scratch.

Dependency hell is a common problem found in software and particularly in HEP software ecosystem. We would like to discuss an improvement of artifact management (“lazy-install”) system as a solution to the “dependency hell” problem. HEP software stack usually consists of multiple sub-projects with dependencies. The development model is often distributed, independent and non-coherent among the sub-projects. We believe that software should be designed to take advantage of other software components that are already available, or have already been designed and implemented for use elsewhere rather than "reinventing the wheel". In our contribution, we will present our approach to artifact management system of ROOT together with a set of examples and use cases.
\end{abstract}

\section{Introduction}

During the different stages of development, through the years ROOT \cite{root} was using various build tools to manage its code, starting from custom build tools going through recently-removed Makefiles and currently using CMake~\cite{cmake} -- a cross-platform build-generator tool.

Programmers love to hate build systems. Following the talks from notorious developer's gatherings and conferences, often build systems and CMake in particular is ridiculed~\cite{moderncmake}. A set of \textit{Modern CMake} tutorials are developed to encourage good practices among the developers' communities. They suggest to start treating CMake as code and think in targets, use properly CMake interface targets either for exports or imports. ROOT is no different, even though the CMake build system is rather new development it can be improved. Using community good practices addresses variety of problems of ROOT build and installation procedures. It is evident that there were a lot of efforts done by the ROOT team to modernize ROOT CMake code and provide a stable and reliable support for ROOT build system~\cite{rootcmake}.

This paper gives some insights of used good practices and what they allow users to do beyond building and shipping ROOT in the standard way.

\section{Background}

Evolving ROOT CMake to improve ROOT installation also allows decreasing build complexity and enables further decoupling of semantically independent components. Component decoupling is essential for embedding ROOT in other ecosystems as it enables more precise configuration of what the ecosystem actually uses. Ideally, ROOT should allow building ROOT component or package on top of already pre-configured or pre-build ROOT. In order to make the ROOT packaging more flexible and less monolithic and to develop a ROOT-aware dependency manager~\cite{pm}.

%Motivation to evolve ROOT CMake build system is based on the idea of providing more straightforward ROOT installations, to improve extensibility, and decrease the complexity of embedding ROOT in other ecosystems. Ideal goals for ROOT modularization project were to enable the possibility to build ROOT component or package on top of already pre-configured or pre-build ROOT, to make the ROOT packaging more flexible and less monolithic and to develop a ROOT-aware dependency manager. \cite{pm}

Current organisation of key components for ROOT build system is very stable. They use the following, fundamental to CMake, concepts:
\begin{enumerate}
    \item ROOT libraries (library targets) together with its specially organised code and tests;
    \item ROOT build options to manipulate with available to user, set of libraries as build deliverables or artefacts;
    \item a set of standalone projects, integrated in ROOT, such as LLVM/Clang, Clad and etc;
    \item ROOT dependencies divided in two groups: dependencies  hosted by ROOT - \textit{ROOT builtins} and external OS package dependencies.
\end{enumerate}

ROOT CMake build options are divided into the two groups: 10 \% build features, such as \textit{cxx11, cxx14, pch, cling} and 90 \% build options such as \textit{gsl\_shared, xml}. Both groups are interdependent and sometimes outdated. The ROOT options are named after directory hosting a library's sources and CMakeLists.txt. That's why sometimes it looks like ROOT CMake options has an ambiguous naming convention and user doesn't know what will be enabled with the particular option. 

Imagine situation when user is requesting to build ROOT "Core" libraries (libCore, libCling, and libRIO) and to enable machine learning libraries on of them via TMVA option. Outcome will be that instead of enabling only TMVA option, the user will enable other not requested $N$ libraries. This example helps to explain the missing build functionality in ROOT, the possibility of the delayed builds.

As a solution, we propose to introduce the concept of a ROOT sub-package.

\section{Implementation ideas}

\begin{figure}[!h]
\centering
\includegraphics[width=0.75\textwidth]{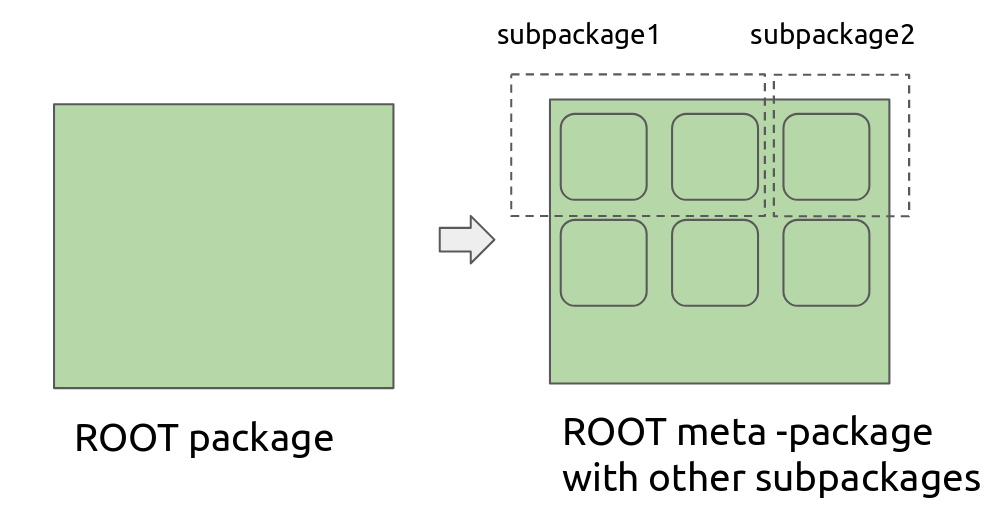}
\caption{Evolution of ROOT package.}
\label{subpackage}
\end{figure}

We can unite libraries with dependencies into the groups or "sub-packages", dealing with ROOT as a "fat" meta-package. A simplified version of this concept is shown on the Figure \ref{subpackage}. In the next subsections, we will try to explain issues that could be resolved in ROOT, introducing ROOT sub-package concepts. Main focus will be to conclude the ROOT layering issues, ROOT dependency management, and options management.

\subsection{Layering ROOT: design goals}

The main intention here is to arrange existing ROOT components into layers. For instance, as a simple ROOT layering example, could be a set of ROOT Math packages. What we would like to achieve is the possibility to enable the next chain:
\begin{equation}
  (core) \rightarrow (mathcore)  \rightarrow (mathmore).  
\end{equation}

Layering concept allows each layer or subpackage to be enabled or disabled independently. It is available via implementation done as an overload of CMake $add\_subdirectory()$ function with iteration loop through special configuration file ROOTPackageMap.cmake, making it similar to a package database. Identical implementation also exists in LLVM project \cite{llvm}: a custom $add\_llvm\_subdirectory()$, $add\_clang\_subdirectory()$ and similar implementations for other LLVM tools \& projects.

\begin{figure}[!h]
\centering
\includegraphics[width=0.4\textwidth]{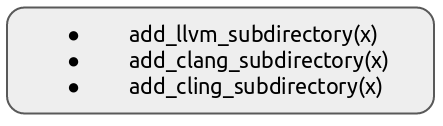}
\caption{Examples of custom $add\_subdirectory()$ from LLVM projects.}
\label{example}
\end{figure}

\begin{figure}[!h]
\centering
\includegraphics[width=0.7\textwidth]{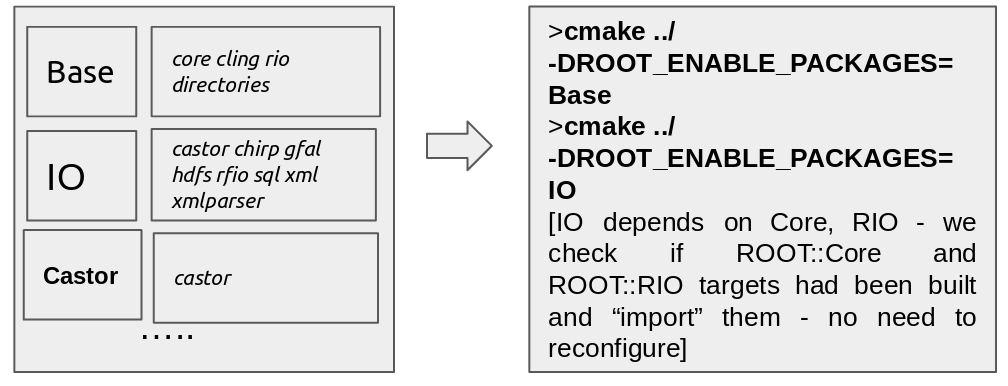}
\caption{ROOT package map and its functionality within ROOT.}
\label{packagemap}
\end{figure}

We propose a new way of organization of ROOT build options, using a map that is similar to a custom database of ROOT packages. Its implementation gives the possibility to enable single ROOT library with its dependencies in one configuration step, which means that we will be able to configure, build, and deliver the ROOT layers iteratively, layer by layer. To enable this feature, we introduces the ROOT package map. On Figure \ref{packagemap} is shown a simplified example of ROOT package map, allowing to enable Base, IO sub-packages together with its dependencies, step by step.

\subsection{Simplification of ROOT dependencies}

\begin{figure}
\centering
\begin{minipage}{\textwidth}
\begin{center}
\subfloat[] {\label{fig:build:a} \includegraphics[width=0.7\textwidth]{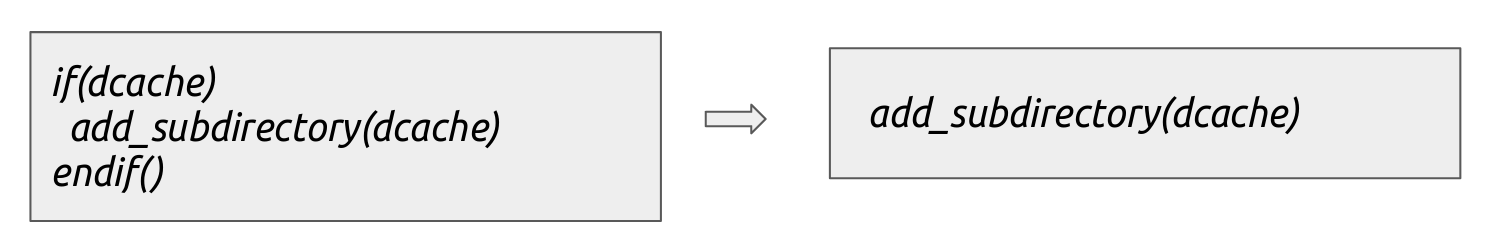}}
\end{center}{}
\end{minipage}\hfill
\begin{minipage}{\textwidth}
\begin{center}
\subfloat[] {\label{fig:build:b} \includegraphics[width=0.5\textwidth]{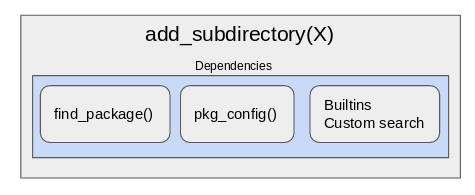}}  
\end{center}{}
\end{minipage}
 \caption{ROOT CMake improvements: (a) ROOT CMake dependency simplification. (b) new procedure for ROOT CMake $add\_subdirectory()$.}
\label{fig:performance1}
\end{figure}

ROOT has a very complex map of dependencies, both internal and external. The idea is to simplify how ROOT dependencies are treated. It will allow to introduce CMake code clarity (see an example on Figure \ref{fig:build:a}) and make ROOT more modular. To enable separability of ROOT layers, inside $add\_subdirectory()$ was introduced a way to treat dependencies in a standalone way: ROOT builtins via separate custom search procedure and external dependencies, using CMake $find\_package()$, $pkg\_config()$ or could be even a simple integration of any other CMake based C++ package manager, such as Conan \cite{conan} (check Figure \ref{fig:build:b}).

\section{Results}

The preliminary results show new possibilities to build ROOT iteratively, package by package. Improvements, proposed in this paper will cover most of the use cases requested by the HEP community to enable a way to configure, build, and use ROOT in the more modular way.

As a consequence, it will help during some critical for users situations such as, when user has already built ROOT from sources and desire to extend its functionality without rebuilding ROOT from scratch or \textit{typical ROOT developer case}, when is actively developed only one of the ROOT component and developer wants to test only this component, without re-configuring all ROOT.

New functionality will help to enable additional range of possibilities for ROOT.
\begin{itemize}
    \item ROOT-aware package manager prototype \textit{root-get}: it will enable the ROOT-aware package management with the root-get prototype. \cite{pm}
    \item \textit{OS package management} (e.g. Fedora, Ubuntu and etc.): it will be easier to generate more granular ROOT packages.
    \item Package, dependency and environment manager \textit{Conda}: it will support a root-minimal package to further improve install times. 
\end{itemize}

A new functionality is expected to be enabled in ROOT 6.20.00 for ROOT C++ modules \cite{cxxmodules} as an experimental option.

\section{Acknowledgments}

This work has been supported by U.S. National Science Foundation grant ACI-1450323.

\end{document}